\documentclass[11pt,letterpaper]{article}
\usepackage[utf8]{inputenc}
\usepackage[USenglish,spanish]{babel}
\usepackage{amsmath}
\usepackage{amsfonts}
\usepackage{amssymb}
\usepackage{graphicx}
\usepackage[left=3cm,right=3cm,top=2cm,bottom=3cm]{geometry}

\usepackage[small]{caption} 

\author{Jorge Pinochet}
\title{\textbf{Stephen Hawking y los agujeros negros cuánticos}}
\begin{document}

\author{Jorge Pinochet$^{*}$\\ \\
 \small{$^{*}$Facultad de Educación,}\\
 \small{\textit{Universidad Alberto Hurtado, Erasmo Escala 1825, Santiago, Chile}. japinochet@gmail.com}\\}

\date{}
\maketitle

\begin{center}\rule{0.9\textwidth}{0.1mm} \end{center}
\begin{abstract}
\noindent Hasta el año 1974, el estudio de los agujeros negros estuvo bajo la hegemonía de la relatividad general de Einstein. Sin embargo, ese mismo año, Stephen Hawking incorporó la teoría cuántica y descubrió que los agujeros negros tienen temperatura, entropía, y se evaporan. Estos tres grandes descubrimientos marcaron el nacimiento de lo que ahora podemos denominar \textit{agujeros negros cuánticos}, en oposición a los \textit{agujeros negros clásicos}, cuya descripción se basa únicamente en la relatividad general. El presente trabajo busca proporcionar una introducción a los tres grandes descubrimientos de Hawking, accesible a estudiantes no graduados de ciencias e ingeniería.\\ \\

\noindent \textbf{Palabras clave}: Agujeros negros, temperatura de Hawking, tiempo de evaporación, entropía de Bekenstein-Hawking, estudiantes no graduados de ciencias e ingeniería.   

\end{abstract}

\selectlanguage{USenglish}

\begin{abstract}
\noindent Until 1974, the study of black holes was under the hegemony of Einstein's general relativity. However, that same year, Stephen Hawking incorporated quantum theory and discovered that black holes have temperature, entropy, and evaporate. These three great discoveries marked the birth of what we can now call \textit{quantum black holes}, as opposed to \textit{classic black holes}, whose description is based only on general relativity. The present work seeks to provide an introduction to the three great discoveries of Hawking, accessible to undergraduate students of science and engineering. \\ \\

\noindent \textbf{Keywords}: Black holes, Hawking temperature, evaporation time, Bekenstein-Hawking entropy, undergraduate physics students.\\ 

\noindent \textbf{PACS}: 01.40.Fk; 04.70.Dy; 01.30.lb.      

\begin{center}\rule{0.9\textwidth}{0.1mm} \end{center}
\end{abstract}

\selectlanguage{spanish}

\maketitle

\section{Introducción}

Los agujeros negros son una de las predicciones más enigmáticas y sorprendentes de la relatividad general, que es la teoría de la gravedad propuesta por Albert Einstein en 1915 para ampliar y perfeccionar la ley de gravitación de Newton [1, 2]. Desde 1916, cuando surgió el primer indicio teórico de la existencia de los agujeros negros, hasta la década de 1964-1974, cuando se produjo la denominada \textit{edad de oro} en el estudio de estos objetos [3], la investigación en el campo estuvo bajo la hegemonía de la relatividad general. Sin embargo, en 1974 un joven y prometedor físico británico, Stephen Hawking, combinó con éxito la relatividad general y la teoría cuántica, y descubrió que los agujeros negros no son tan negros, contradiciendo el conocimiento acumulado desde 1915. En concreto, Hawking demostró que los agujeros negros emiten un tipo de radiación térmica conocida desde entonces como \textit{radiación de Hawking}, cuya existencia tiene tres grandes implicancias físicas: los agujeros negros tienen una temperatura no nula, tienen entropía, y reducen su masa en forma gradual en un proceso llamado \textit{evaporación} [4, 5]. Estos tres importantes resultados han permitido trazar una línea divisoria entre lo que ahora podemos denominar \textit{agujeros negros clásicos}, cuya descripción se basa únicamente en la relatividad general, y los \textit{agujeros negros cuánticos}, cuyo estudio combina la relatividad general y la teoría cuántica.\\

Dada la importancia de los descubrimientos del físico británico, a través de los años se ha acumulado una abundante literatura sobre el tema. Sin embargo, salvo contadas excepciones [6, 7], los textos que analizan las contribuciones de Hawking al estudio de los agujeros negros pueden dividirse en dos categorías, cuyos niveles de dificultad son diametralmente opuestos. Por una parte está la literatura técnica, destinada a personas con un gran dominio de la física y las matemáticas. Por otra parte está la divulgación, destinada a personas sin formación científica. Por tanto, existe un evidente vacío de literatura de nivel intermedio, cuya dificultad se encuentre a medio camino entre la divulgación y la literatura técnica. El propósito del presente trabajo es presentar una introducción accesible y actualizada a los tres grandes descubrimientos del recientemente fallecido genio británico. El artículo está destinado principalmente a aquellos lectores cuya formación en física y matemáticas los sitúa entre el especialista y el lego (aunque más cerca de este último). Dentro de este segmento se encuentran, por ejemplo, los estudiantes de pregrado de carreras del ámbito de las ciencias y la ingeniería. Desde esta perspectiva, el artículo puede resultar provechoso como material de análisis en un curso de pregrado de astronomía o física moderna.\\

En las secciones 2 y 3 se ofrece una visión panorámica del tema, donde se introducen los conceptos de agujero negro clásico y radiación de Hawking. En la sección 4 se analizan las ecuaciones que sintetizan los tres grandes descubrimientos de Hawking: la ecuación para la temperatura, el tiempo de evaporación y la entropía de un agujero negro. En la sección 5 se presentan derivaciones heurísticas de estas ecuaciones, basadas en argumentos físicos muy simples. Para finalizar, en la sección 6 se examina la posibilidad de detectar agujeros negros cuánticos.

\section{¿Qué es un agujero negro?}
De acuerdo con un importante resultado de la relatividad general conocido como \textit{teorema de no pelo} [8], existen únicamente tres variables clásicas observables externamente que definen un agujero negro: masa, carga eléctrica y momentum angular (rotación)\footnote{Básicamente, el teorema de no pelo asegura que un agujero negro no tiene <<memoria>> respecto de las propiedades o <<pelos>> del objeto a partir del cual se formó. No importa la tamperatura, la forma, la composición química o el campo magnético de dicho objeto; tampoco importa si el objeto es una estrella, un planeta, una mota de polvo o una partícula subatómica. Las únicas propiedades que un agujero negro retiene son la masa, la carga eléctrica y al momentum angular; luego, si dos agujeros negros tienen los mismos valores para estas tres propiedades, no es posible distinguir uno de otro, ya que carecen de pelos que permitan identificarlos.}. Por consiguiente, el agujero negro más simple es el que carece de rotación y es eléctricamente neutro. A este objeto se le conoce como \textit{agujero negro estático}, y su descripción matemática solo depende de la masa. Para simplificar la exposición, en adelante centraremos la atención en este tipo de agujero. Ello no supondrá perdida de generalidad, puesto que, como veremos más adelante, en lo fundamental los tres descubrimientos de Hawking se aplican a los agujeros estáticos. \\

Un agujero negro se forma cuando se produce una alta concentración de masa, o su equivalente en energía, en una región esférica cerrada del espacio denominada \textit{horizonte}. La relatividad general asegura que el campo gravitacional dentro del horizonte es tan intenso, que la luz no puede escapar de su interior y queda atrapada para siempre\footnote{Fue esta capacidad de “atrapar la luz” lo que llevó al físico teórico John Archibald Wheeler a acuñar el término \textit{agujero negro} en 1967.}. Pero como según la física relativista nada en el universo puede desplazarse más rápido que la luz, se concluye que ninguna forma de materia o energía encerrada en el horizonte puede atravesarlo hacia el exterior. \\

\begin{figure}
  \centering
    \includegraphics[width=0.3\textwidth]{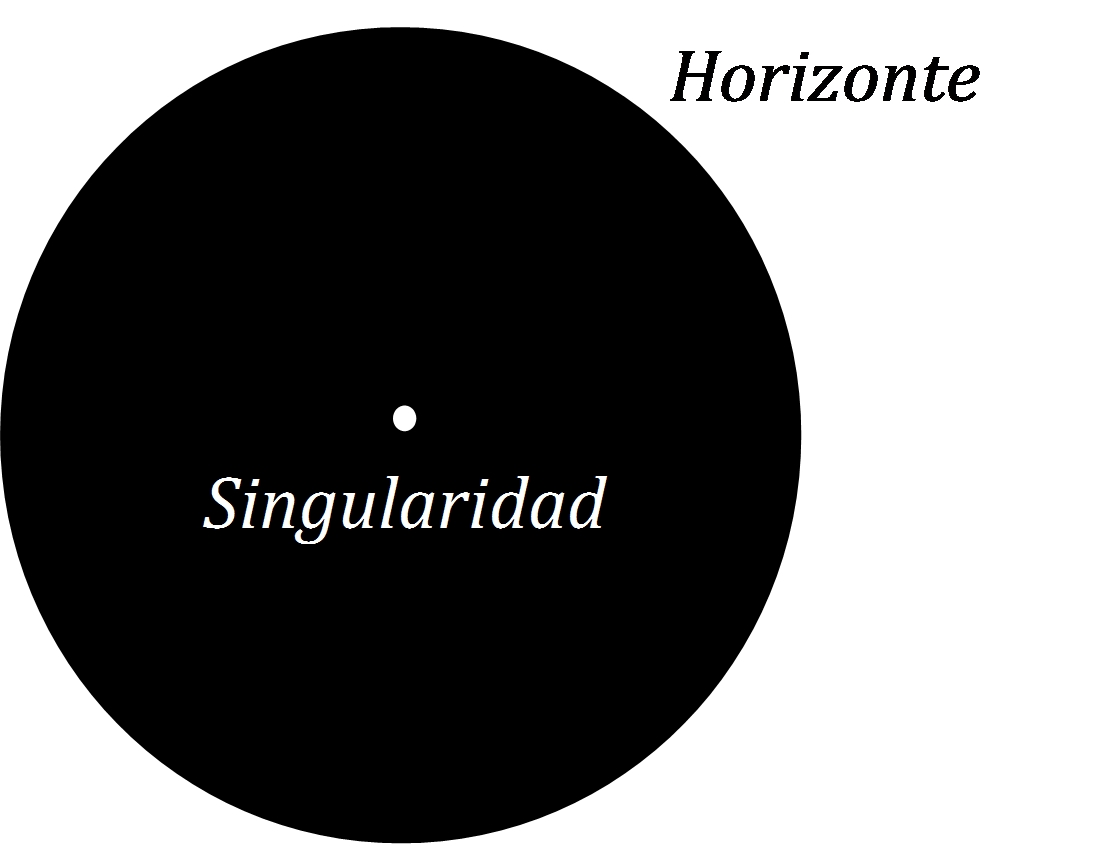}
  \caption{Un agujero negro estático tiene una singularidad central ($S$) rodeada por un horizonte esférico cerrado ($H$).}
\end{figure}

Si designamos como $M_{BH}$ a la masa del agujero negro (Black Hole), el radio que caracteriza al horizonte puede calcularse mediante una ecuación encontrada en 1916 por el astrónomo alemán Karl Schwarzschild\footnote{Existe un argumento newtoniano intuitivo para obtener la ecuación (1). La velocidad de escape desde la superficie de un objeto esférico masivo de radio $R$ y masa $M$ es $V_{e} = (2GM/R)^{1/2}$. Tomando $V_{e} = c$ y resolviendo para $R$ se obtiene: $R = 2GM/c^{2}$. El significado físico de esta expresión es claro: ninguna forma de materia o energía contenida dentro de la superficie esférica limitada por $R$ puede escapar, ya que para ello necesitaría una rapidez mayor que $c$.} [9, 10]:

\begin{equation} 
R_{S} = \frac{2GM_{BH}}{c^{2}} \approx 3.00\times 10^{3}m \left( \frac{M_{BH}}{M_{\odot}}\right), 
\end{equation}

donde $R_{S}$ se conoce con \textit{radio de Schwarzschild}\footnote{En el marco intuitivo de la gravitación newtoniana, $R_{S}$ puede imaginarse como la distancia que separa el centro del agujero negro de su horizote; no obstante, en relatividad general esta interpretación es incorrecta, ya que $R_{S}$ se considera una coordenada, y no una distancia física} , $G = 6.67 \times 10^{-11} N\cdot m^{2} \cdot kg^{-2}$ es la constante de gravitación universal, y $c = 3,00 \times 10^{8} m \cdot s^{-1}$ es la rapidez de la luz en el vacío. Siguiendo una práctica habitual en astrofísica, $R_{S}$ se ha expresado en unidades de masa solar, $M_{\odot} = 1.99 \times 10^{30}kg$, y la ecuación (1) ha quedado expresada en metros ($m$). Para tener una idea de las colosales concentraciones de materia implicadas en la formación de un agujero negro, notemos que según esta ecuación, para convertir al Sol en un agujero negro, sería necesario comprimirlo hasta que su radio fuera de al menos $3 km$, lo que equivale a una cienmilésima parte del radio solar.\\ 

Según la relatividad general, si una determinada cantidad de masa, o su equivalente en energía, se comprime dentro del horizonte, se inicia un colapso gravitatorio que no puede detenerse. El colapso culmina cuando la masa-energía queda reducida a un punto matemático de densidad infinita denominado singularidad, localizado en el centro del horizonte (ver Figura 1). Aunque el horizonte no es una superficie física, puede visualizare como una membrana unidireccional que solo permite el flujo de materia o energía hacia el interior [11]. Como un agujero negro no puede emitir ninguna forma de radiación desde su horizonte, las leyes de la termodinámica aseguran que su temperatura debe ser estrictamente nula, pues de otro modo el agujero negro emitiría radiación térmica y no sería negro. No obstante, como se verá en la siguiente sección, los agujeros negros no son tan negros, ya que la teoría cuántica predice que deben emitir radiación térmica y por lo tanto es posible asociar temperatura y entropía al horizonte.

\section{El principio de incertidumbre y la radiación de Hawking} 
Una de las consecuencias más sorprendentes de la unión entre la física relativista y la teoría cuántica es que el espacio nunca puede estar completamente vacío de energía, o de su equivalente en masa. Este fenómeno no tiene paralelo en la física clásica, y es consecuencia del principio de incertidumbre de Heisenberg. En una de sus formulaciones, este principio establece que si $\Delta E$ es la incertidumbre en la medida de la energía en un determinado volumen de espacio, y si $\Delta t$ es la incertidumbre en el tiempo que dura dicha medida, el valor mínimo que puede tomar el producto de estas cantidades es: 

\begin{equation} 
 \Delta t \Delta E = \frac{\hbar}{2} \approx \hbar, 
 \end{equation} 

donde $\hbar = h/ 2\pi = 1.05 \times 10^{-34} J\cdot s$ es la constante de Planck reducida. Por otra parte, la equivalencia relativista entre masa y energía permite relacionar la incerteza en la masa $\Delta m$ con la incerteza en la energía mediante la expresión $\Delta E = \Delta mc^{2}$. A partir de esta relación, la ecuación (2) puede escribirse como:

\begin{equation} 
 \Delta t \Delta m = \frac{\hbar}{2c^{2}} \approx \frac{\hbar}{c^{2}}. 
 \end{equation} 

Estas ecuaciones revelan que si la masa-energía de una determinada región del espacio vacío fuera exactamente cero en todo instante, se tendría que $\Delta E = \Delta m=0$, y por tanto $\hbar = 0$. Como esto es una contradicción, se concluye que la masa-energía del vacío no pude ser nula, de modo que debe existir un valor residual. Debido a la naturaleza aleatoria de los procesos cuánticos, esta masa-energía residual no tiene un valor fijo, sino que sufre rápidas variaciones temporales llamadas \textit{fluctuaciones del vacío}, que se manifiestan como pares partícula-antipartícula que aparecen súbitamente, se separan y luego vuelven a juntarse para aniquilarse mutuamente, siendo reabsorbidas por el vacío, lo cual significa que desaparecen completamente (ver figura 2)\\

Las fluctuaciones del vacío duran tan poco tiempo, que el proceso nunca puede observarse en forma directa, y los pares partícula-antipartícula desaparecen antes de que puedan detectarse. Por esta razón, estas partículas se denominan \textit{virtuales}. El lapso de vida característico de las partículas virtuales puede calcularse a partir de las ecuaciones (2) y (3):

\begin{equation} 
\Delta t \approx \frac{\hbar}{\Delta E} \approx \frac{\hbar}{c^{2} \Delta m}. 
\end{equation}

Una consecuencia muy importante de la imposibilidad de observar partículas virtuales, es que las fluctuaciones del vacío mantienen a resguardo la ley de conservación de la energía, ya que nunca se produce una violación observable de esta ley. De hecho, se puede demostrar que las fluctuaciones respetan las leyes de conservación, tanto de cantidades continuas (energía, momentum angular, momentum lineal) como discretas (carga eléctrica, número leptónico, número bariónico, etc.).\\

\begin{figure}
  \centering
    \includegraphics[width=0.4\textwidth]{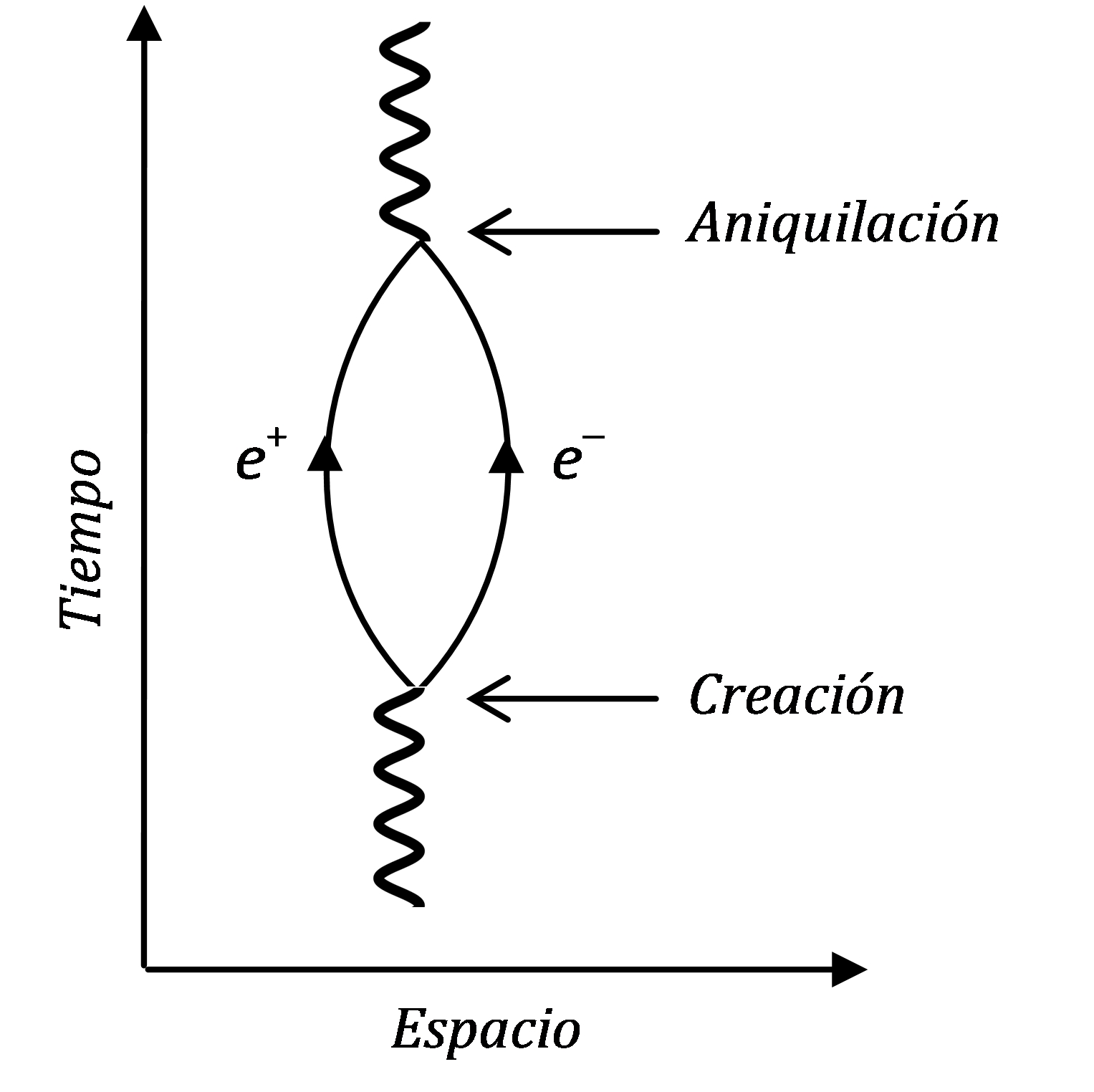}
  \caption{Un par de partículas virtuales, compuestas por un electrón $(e^{-})$ y un positrón $(e^{+})$ son creados en un determinado instante. En la medida que transcurre el tiempo, las partículas se separan cierta distancia para luego volver a encontrarse, aniquilarse, y ser reabsorbidas por el vacío.}
\end{figure}

¿Cómo se relaciona este fenómeno con la radiación y la temperatura de Hawking? La imagen clásica presentada por Hawking [11, 12], que según algunos especialistas no es del todo correcta [13], pero que  a juicio del autor resulta muy útil pedagógicamente, es la siguente: como todo volumen de espacio está sujeto a fluctuaciones cuánticas, éstas también tendrán lugar en la región que se encuentra justo fuera del horizonte. Debido a la gran intensidad del campo gravitacional en esa región, las fuerzas de marea pueden separar definitivamente una partícula de su correspondiente antipartícula. En estas condiciones, una partícula será absorbida por el agujero, y la otra podrá escapar, convertida en una partícula real, porque no podrá volver a reunirse con su compañera para aniquilarse mutuamente. Por lo tanto, la energía que permite a esta partícula escapar, es tomada directamente del campo gravitacional del agujero negro.\\

Para un observador externo, la partícula que escapa parecerá haber sido emitida desde el horizonte [12]. La acción sostenida de este proceso a través del tiempo, sobre un gran número de pares partícula-antipartícula, produce un flujo continuo emitido en todas direcciones, denominado \textit{radiación de Hawking}. Este flujo tiene un espectro térmico, y permite asociar al horizonte una temperatura característica, conocida como \textit{temperatura de Hawking}. Cada partícula de energía positiva $+E$ que es emitida, tiene una compañera de energía negativa $-E$ que al ser absorbida por el agujero negro, reduce la masa de éste, evitando que se viole la ley de conservación de la masa-energía [12]. Este mecanismo conduce a la evaporación gradual del agujero negro. Además, de acuerdo con las leyes de la termodinámica, si el agujero negro tiene temperatura, entonces debe tener entropía. En la siguiente sección se analizan con más detalle las implicancias físicas de la radiación de Hawking.

\begin{figure}
  \centering
    \includegraphics[width=0.4\textwidth]{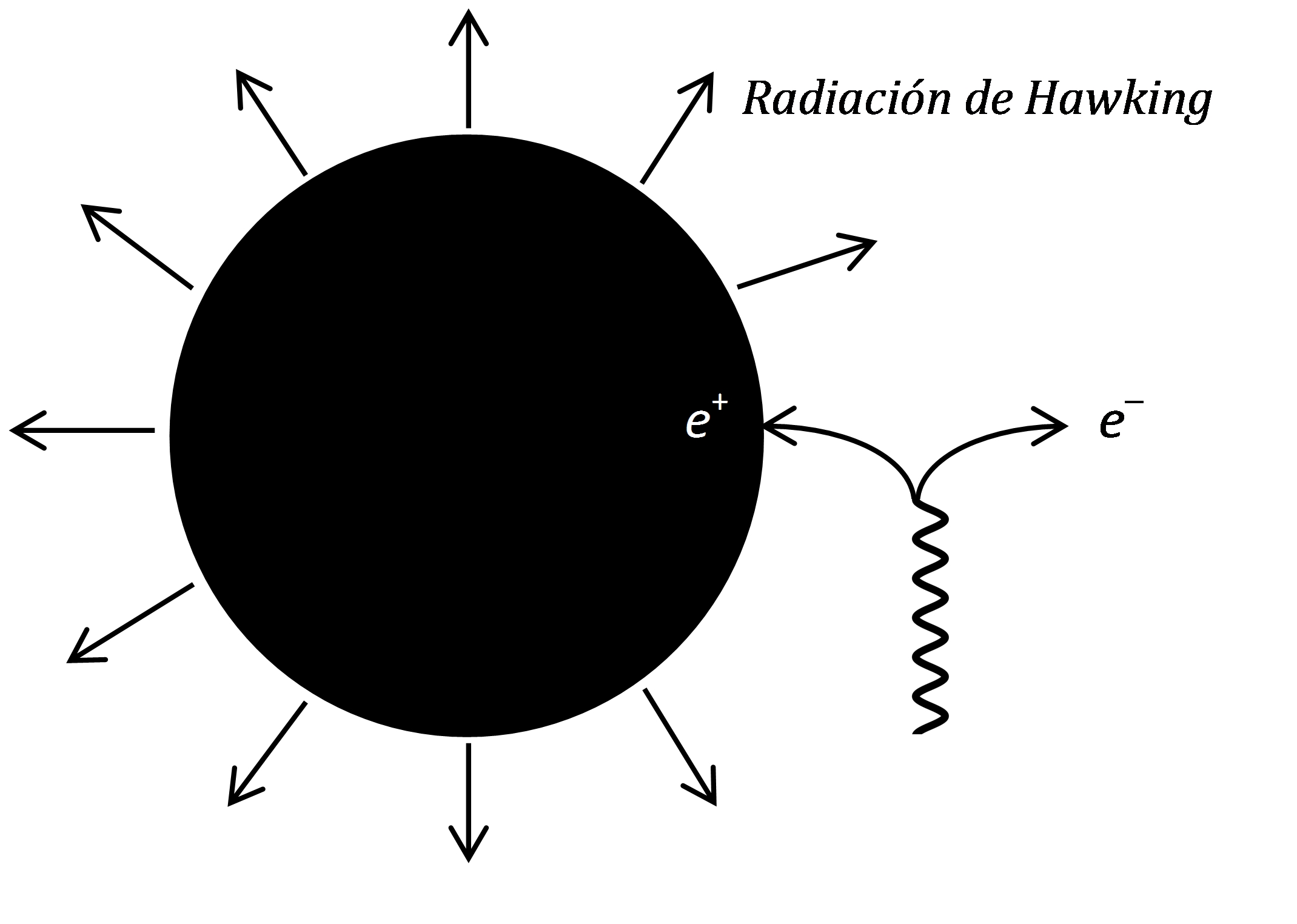}
  \caption{Cuando se produce un par partícula-antipartícula fuera del horizonte de un agujero negro, una partícula ($e^{+}$) es atrapada por la gravedad y cae dentro del horizonte. La otra partícula ($e^{-}$) logra escapar. Externamente, esta partícula es detectada como radiación de Hawking con temperatura $T_{H}$.}
\end{figure}

\section{Los tres grandes descubrimientos de Hawking}

A principios de la década de 1970, el físico teórico griego Demetrios Christodoulou y el propio Hawking obtuvieron importantes resultados que demostraban la imposibilidad de reducir la masa-energía de un agujero negro estático [14-16]. Estos resultados, obtenidos en el marco de la relatividad general clásica, son consecuencia directa del hecho que ninguna forma de materia o energía atrapada por un agujero negro, puede atravesar el horizonte hacia el exterior. Sin embargo, los hallazgos de Christodoulou y Hawking no prohíben la reducción de masa-energía en agujeros negros con carga eléctrica y/o momentum angular. De hecho, en 1965, el físico y matemático británico Roger Penrose propuso un mecanismo físico que permite extraer energía de un agujero negro rotatorio [17]. \\

Fue en este escenario donde Hawking propuso la revolucionaria idea de que un agujero negro estático en el vacío emite radiación térmica, es decir, radiación de Hawking. De acuerdo con las leyes de la termodinámica, esto implica que un agujero negro estático tiene temperatura, entropía y se evapora. El hecho que estos tres resultados se apliquen a agujeros en el vacío es un aspecto importante del hallazgo de Hawking, pues significa que la emisión de radiación no depende de mecanismos relacionados con la presecia de material en las afueras del horizonte, como sucede con la acreción, que es un proceso físico que genera grandes emisiones de radiación en agujeros negros que forman parte, por ejemplo, de sistemas binarios. Al contrario, la radiación de Hawking proviene directamente del agujero negro. En este contexto no es de extrañar que en su momento, los resultados de Hawking causaran revuelo, ya que desafiaban abiertamente a la relatividad general, y contradecían la definición misma de agujero negro. Estos descubrimientos fueron publicados por Hawking en un breve artículo que apareció en la revista \textit{Nature} en 1974 [4], y luego fueron ampliados en un trabajo más técnico publicado en la revista \textit{Communications in Mathematical Physics} [5]. A continuación vamos a analizar detenidamente estos tres grandes resultados, comenzando con la temperatura asociada al horizonte.\\

\subsection{Temperatura de Hawking}

De acuerdo con lós cálculos de Hawking, el horizonte emite radiación térmica como un cuerpo negro con una temperatura absoluta dada por:

\begin{equation} 
 T_{H} = \frac{\hbar c^{3}}{8\pi kGM_{BH}} = 6.17 \times 10^{-8} K \left( \frac{M_{\odot}}{M_{BH}}\right). 
 \end{equation} 

Esta es la \textit{temperatura de Hawking}, donde $k = 1.38 \times 10^{-23} J \cdot K^{-1}$ es la constante de Boltzmann, y nuevamente se han usado unidades de masa solar, quedando $T_{H}$ expresada en kelvin ($K$). Un punto importante pero sutil respecto de este resultado, sobre el cual no podemos ahodar, es que $T_{H}$ no es la temperatura en las inmediaciones del horizonte, sino que es la temperatura medida a una gran distancia (idealmente infinita).\\

La proporcionalidad inversa entre $T_{H}$ y $M_{BH}$ sugiere que $T_{H}$ solo es significativa para agujeros negros comparativamente pequeños y poco masivos. En efecto, en su artículo seminal de 1974, Hawking especuló con la posible existencia de agujeros negros pequeños, cuyas masas son menores que $\sim 10^{12 } kg$. De acuerdo con la ecuación (1), esta cifra corresponde a un radio de Schwarzschild menor que $\sim 10^{-15 }m$, lo que equivale al tamaño típico de un núcleo atómico. Tomando $M_{BH}\sim 10^{12} kg$ en la ecuación (5) se encuentra que $T_{H} \sim 10^{10} K$, una temperatura muy elevada, y que supera en un factor $\sim 10^{3}$ a la temperatura en el centro de una estrella típica, que es $\sim 10^{7}K$. Por otra parte, se observa que para los denominados agujeros negros estelares, cuyas masas son del orden de la masa solar, $T_{H}$ es sumamente baja, ya que si tomamos $M_{BH} \sim M_{\odot}$, se obtiene $T_{H} \approx 10^{-8}K$, una temperatura cercana al cero absoluto, y que es indetectable mediante observaciones astronómicas [18]. En la última sección volveremos sobre este punto, donde discutiremos la posibilidad de verificar observacionalmente los descubrimientos de Hawking.\\

¿Porqué $T_{H}$ es inversamente proporcional a $M_{BH}$? Para responder esta pregunta debemos recordar que el campo gravitacional newtoniano $a$ en la superficie de un objeto esférico de masa $M$ y radio $R$ viene dado por: $a = GM/R^{2}$. Si en esta ecuación tomamos $R = R_{S}$, donde $R_{S}$ está dado por la ecuación (1), se obtiene el campo gravitacional en el horizonte:

\begin{equation} 
a = \frac{c^{4}}{4GM_{BH}}. 
\end{equation}

Aunque utilizamos la física de newton para obtener esta ecuación, el resultado coincide exactamente con la expresión relativista para la denominada \textit{gravedad superficiel en el horizonte}\footnote{Para ser precisos, en relatividad general la aceleración cerca el horizonte es $a=(GM_{BH}/R^{2})/(1-R_{S}/R)^{1/2}$; para $R = R_{S}$ se observa que $a$ se vuelve infinita. Como no es conveniente trabajar con cantidades infinitas, se define la \textit{gravedad superficial en el horizonte}, que es el numerador de la ecuación anterior evaluada en $R_{S}$.}. Si combinamos las ecuaciones (5) y (6) resulta: 

\begin{equation} 
T_{H} = \frac{\hbar}{2\pi ck}a. 
\end{equation}

Por lo tanto, preguntar por qué $T_{H}$ es inversamente proporcional a $M_{BH}$ equivale a preguntar por qué es directamente proporcional a $a$. La respuesta ha sido sugerida antes. Las partículas de la radiación de Hawking obtienen su energía a expensas del campo gravitacional en el horizonte, de modo que un aumento de $a$ conlleva un incremento de la energía. Y como en general, la temperatura depende de la energía de la radiación térmica, se concluye que $T_{H}$ debe ser proporcional a $a$. La ecuación (7) también revela que cuanto mayor es $a$, mayor es el número de partículas que son emitidas desde el horizonte (mayor intensidad de radiación), pues dicho número depende de la cantidad de partículas virtuales que se convierten en reales, lo que a su vez depende de la magnitud de las fuerzas de marea, que son proporcionales a $a$. \\

\subsection{Tiempo de evaporación}

Como sucede con cualquier cuerpo caliente, un agujero negro emite radiación térmica compuesta preferentemente por fotones. La radiación de Hawking se lleva una parte de la masa-energía del agujero negro, generando una reducción gradual de $M_{BH}$ que da lugar al proceso de evaporación, que fue el segundo gran resultado obtenido por Hawking en 1974-1975. Si $T_{H}$ es lo suficientemente elevada, o $M_{BH}$ es lo suficientemente baja, es posible que además de fotones, la radiación de Hawking esté compuesta por partículas masivas, tales como neutrinos, electrones, protones, etc. En efecto, a partir de la ecuación (5) vemos que la energía característica de una partícula de la radiación de Hawking es $kT_{H}$, de modo que en la práctica una partícula de energía $E$ o masa $E/c^{2}$ solo podrá ser emitida si:

\begin{equation} 
E\lesssim kT_{H} = \dfrac{\hbar c^{3}}{8 \pi GM_{BH}} \sim 10^{-12} eV \left(\dfrac{M_{\odot}}{M_{BH}} \right), 
\end{equation}

donde $1 eV = 1.60 \times 10^{-19} J$. Esta ecuación revela que en la medida que el agujero se evapora y $M_{BH}$ disminuye, la masa-energía de las partículas emitidas es cada vez mayor, pudiendo cubrir toda la gama de partículas existentes. Por ejemplo, si comenzamos con un agujero negro de masa solar, $M_{BH} = M_{\odot} = 1.99\times 10^{30} kg$ ($R_{S} \sim 10^{3} m$), la energía máxima de las partículas emitidas es $E = 10^{-12} eV$, que corresponde a fotones de radioondas. Después de un largo proceso de evaporación (más adelante veremos qué tan largo es este proceso), el agujero negro habrá reducido su masa, por ejemplo, hasta $M_{BH} = 10^{10} kg \sim 10^{-20} M_{\odot}$ ($R_{S} \sim 10^{-17} m$). Ahora, la energía máxima de las partículas emitidas es $E = 10^{8} eV$, lo que equivale a una masa máxima $m = 10^{8} eV/c^{2}$; estas cifras son del orden de la masa-energía de un protón.\\  

Si llevamos la evaporación hasta sus últimas consecuencias, de acuerdo con la ecuación (5) $T_{H}$ debiera volverse infinita en los últimos instantes de vida del agujero negro, cuando $M_{BH}$ tiende a cero. Este resultado sugiere que la ecuación (5) no tiene validez general. La solución a este problema es uno de los grandes desafíos de la física teórica, y aún no encuentra una solución satisfactoria, aunque se han realizado progresos [19]. Sin embargo, por razones de consistencia con otras teorías físicas que han sido ampliamente confirmadas, los especialistas coinciden en que la ecuación (5) es correcta, siempre que $M_{BH}$ no sea demasiado pequeña.  \\

El tiempo que un agujero negro en el vacío demora en reducir su masa a su valor mínimo\footnote{Argumentos teóricos sugieren que este valor mínimo es extremadamente pequeño pero distinto de cero.}, se denomina \textit{tiempo de evaporación}, $t_{ev}$, y viene dado por:

\begin{equation} 
t_{ev} \approx \frac{G^{2} M_{BH}^{3}}{c^{4}\hbar} \sim 10^{70}s \left(\frac{M_{BH}}{M_{\odot}}\right)^{3}, 
\end{equation}

donde $M_{BH}$ es la masa inicial del agujero negro. Este fue el resultado encontrado en 1974 por Hawking\footnote{Para ser exactos, la estimación efectuada por Hawking es $t_{ev} \sim 10^{71}s \left(\frac{M_{BH}}{M_{\odot}}\right)^{3}$. La diferencia de un factor 10 entre esta expresión y la relación (9) se debe a las aproximaciones numéricas empleadas.}. Sin embargo, se trata de una estimación cruda de $t_{ev}$. Un cálculo más detallado fue desarrollado en 1976 por el físico canadiense Don Page [20], quien encontró la siguiente expresión:

\begin{equation} 
t_{ev} \approx 10^{73}s \left(\frac{M_{BH}}{M_{\odot}}\right)^{3}.  
\end{equation}

Notemos que esta ecuación difiere en un factor $10^{3}$ de la estimación original de Hawking. El hecho que $t_{ev}$ sea proporcional al cubo de la masa inicial revela que para agujeros negros estelares, $t_{ev}$ es extraordinariamente grande, pues para $M_{BH} = M_{\odot}$, se encuentra que $t_{ev} \sim 10^{73}s$; esta cifra es 56 órdenes de magnitud mayor que la edad del universo, que es de $\sim 10^{17}s$. Por tanto, tal como sucedió con la temperatura de Hawking, los cálculos muestran claramente que para agujeros negros estelares, los efectos cuánticos predichos por Hawking son inobservables. \\

No obstante, incluso suponiendo que los astrónomos dispusieran de un tiempo infinito para observar la evaporación de agujeros estelares, existe otra dificultad que debemos considerar. Como la radiación de fondo de microondas que llena el universo tiene una temperatura $T_{RFM }= 2.73K$, que es mucho mayor que la temperatura típica de un agujero estelar, $T_{H} \sim 10^{-8} K$, debe existir un flujo neto de radiación dirigido hacia el interior del horizonte. Esto implica un incremento de $M_{BH}$ y una reducción de $T_{H}$, y por lo tanto los agujeros negros estelares no se están evaporando. Si como indica la evidencia observacional, el universo está destinado a expandirse para siempre, llegará un momento en que $T_{RFM} < T_{H}$, y entonces la masa de los agujeros estelares comenzará a reducirse. Sin embargo, para que esto suceda, el universo tendría que expandirse al menos un factor $\sim 10^{8}$  respecto de su tamaño actual, de modo que no existe ninguna posibilidad de observar la evaporación de agujeros estelares\footnote{Se puede demostrar que $T_{RFM}$ es inversamente proporcional al denominado \textit{factor de escala}, $R$, que proporciona una medida (adimensional) del tamaño del universo. La relación matemática puede escribirse como $T_{RFM1}/T_{RFM2} =R_{2}/R_{1}$. Si $T_{RFM1} = 2.73 K$, y $T_{RFM2} \sim 10^{-8}K$, entonces $R_{2}/R_{1}\sim 10^{8}$}. \\

\subsection{Entropía}

Como la temperatura implica la existencia de entropía, Hawking concluyó que un agujero negro debe tener entropía. Este fue el tercer gran resultado obtenido por el físico británico en 1974-1975. De acuerdo con los cálculos de Hawking, la entropia $S_{BH}$ de un agujero negro es proporcional al área $A$ del horizonte:

\begin{equation} 
S_{BH} = \frac{kc^{3}A}{4 \hbar G}, 
\end{equation}

donde $A$ se calcula a partir de la ecuación (1) [10]:

\begin{equation} 
 A = 4\pi R_{S}^{2} = \frac{16 \pi G^{2}M_{BH}^{2}}{c^{4}}. 
 \end{equation} 

El subíndice $BH$ en la ecuación (11) corresponde a las iniciales de Bekenstein-Hawking (aunque también podría interpretarse como las iniciales de Black-Hole) ya que la noción de entropía de agujero negro fue sugerida originalmente en 1973 por el físico teórico mexicano-estadounidense-israelí Jacob Bekenstein [21], y luego fue precisada por Hawking. De hecho, los cálculos de Bekenstein solo le permitieron hacer una estimación gruesa de la entropía de un agujero negro, pero fue Hawking quien efectuó un cálculo detallado y comprendió con mayor profundidad las implicancias físicas de dicha entropía. \\

La ecuación (11) puede reescribirse de una forma sugerente empleando la \textit{longitud de Planck}:

\begin{equation} 
l_{P} \equiv \left( \frac{\hbar G}{c^{3}} \right)^{1/2} = 1.61 \times 10^{-35} m. 
\end{equation}

Esta expresión fue introducida por el físico alemán Max Planck en 1899 como parte de un sistema de unidades universal basado en las constantes fundamentales de la naturaleza: $\hbar$, $G$, y $c$. Es fácil advertir que $l_{p}$ es extraordinariamente pequeña. De hecho, el radio típico de un núcleo atómico es $\sim 10^{-15}m$, lo que es un factor $\sim 10^{20}$ mayor que $l_{P}$. De acuerdo con la interpretación moderna, $l_{P}$ corresponde a la escala de tamaño más pequeña. En otras palabras, $l_{P}$ es la unidad de longitud mínima a la que se le puede atribuir un significado físico [22]. Para escribir $S_{BH}$ en función de $l_{P}$, reordenamos términos en la ecuación (11):

\begin{equation} 
S_{BH} = \frac{kA}{4\left( \hbar G/c^{3}\right)} = \frac{kA}{4l_{P}^{2}}. 
\end{equation}

A partir de esta ecuación, podemos interpretar $S_{BH}$ como una medida del número de unidades de área mínima $l_{P}^{2}$ contenidas en el horizonte. Como $l_{P}^{2} \sim 10^{-70}m^{2}$  -un área inimaginablemente pequeña-, el cociente $A/l_{P}^{2}$ será en general muy grande. Esto sugiere que la entropía típica de un agujero negro debe ser extraordinariamente elevada. Un poco más adelante vamos a estimar este resultado mediante un cálculo directo. \\

Para verificar que esta ecuación no viola la segunda ley de la termodinámica, según la cual la entropía de un sistema cerrado nunca disminuye, se define la entropía generalizada: 

\begin{equation}
S_{G} = S + S_{BH} = S + \frac{kA}{4l_{P}^{2}},
\end{equation}

donde $S$ representa la entropía total de toda la materia del universo, fuera de los agujeros negros. Se observa que aunque $S_{BH}$ y $A$ puedan aumentar/decrecer individualmente, $S_{G}$ nunca decrecerá; si el agujero negro absorbe material de su entorno, $M_{BH}$ aumenta y por tanto $A$ aumenta, pero al mismo tiempo $S$ disminuye, de modo que $S_{G}$ no decrece. Por otra parte, la emisión de radiación de Hawking disminuye $M_{BH}$ y $A$, pero aumenta $S$, de modo que nuevamente $S_{G}$ no decrece.\\ 

Para seguir desentrañando el significado de la ecuación (11), recordemos que en mecánica estadística se define la entropía como el logaritmo del número $W$ de microestados de un sistema físico que son compatibles con un determinado macroestado; en otras palabras, la entropía es proporcional al número $W$ de configuraciones que los componentes microscópicos de un sistema pueden adoptar, de manera que el sistema se vea idéntico macroscópicamente. Por lo tanto, podemos estimar la entropía $S_{BH}$ de un agujero negro como:

\begin{equation} 
S_{BH} = k \ln W_{BH}. 
\end{equation}

Eliminando $S_{BH}$ de la ecuaciones (11) y (16):

\begin{equation} 
W_{BH} = e^{c^{3}A/4\hbar G} = e^{4\pi GM_{BH}^{2} /\hbar c}. 
\end{equation}

Como el macroestado de un agujero estático solo depende de $M_{BH}$, si nos apegamos a la definición de entropía dada antes [5, 21, 23], se concluye que $W_{BH}$ es el número de microestados de un agujero estático que son compatibles con un valor determinado de $M_{BH}$. Por ejemplo, si en la ecuación (17) introducimos el valor de $M_{BH}$ para un agujero negro estelar, $\sim 10^{30}kg$ se encuentra que $S_{BH} \sim 10^{10^{77}}$, un cifra  colosal que sugiere que los agujeros negros se cuentan entre los objetos más entrópicos el universo. El físico teórico estadounidense Kip Thorne ha sugerido que $W_{BH}$ es una medida del número de maneras en que podría haberse formado un agujero negro a partir de una determinada catidad de materia. Un análisis amplio y accesible de este tema se encuentra en un libro de divulgación escrito por el propio Thorne [3]. Sin embargo, la propuesta de Thorne deja muchas preguntas sin respomder ya que hasta el momento no existe ninguna explicación detallada del significado físico  de $W_{BH}$ en términos de algún tipo de configuraciones microscópicas internas de un agujero negro.

\section{Una estimación de los resultados de Hawking}

Las ecuaciones centrales que sintetizan los tres grandes descubrimientos del físico británico son la ecuación (5) para la temperatura de Hawking, la ecuación (9) para el tiempo de evaporación, y la ecuación (11) para la entropía de Bekenstein-Hawking. Para llegar a estos resultados, Hawking elaboró un complejo razonamiento que combina la relatividad general, la teoría cuántica y la termodinámica\footnote{En rigor, la termodinámica está incluida dentro de la mecánica cuántica a través de las estadísticas cuánticas. Sin embargo, con fines pedagógicos, es más claro separar la mecánica cuántica de la termodinámica.}, lo que queda de manifiesto por la presencia en las ecuaciones (5), (9) y (11) de las constantes fundamentales que caracterizan a estas teorías: $G$, $c$, $\hbar$ y $k$. Siguiendo a Hawking, pero a un nivel elemental, a continuación derivaremos heurísticamente estas tres ecuaciones, utilizando conceptos básicos de relatividad, teoría cuántica y termodinámica. Como en las derivaciones heurísticas las contantes adimensionales son poco confiables, las omitiremos de los cálculos.\\

Comencemos estimando $T_{H}$. De acuerdo con la ecuación (5), cuanto menor es $M_{BH}$, mayor es $T_{H}$, mayor es la energía de los fotones emitidos, y menor su longitud de onda. Esto muestra que la longitud de onda $\lambda$ de la radiación de Hawking es proporcional a $M_{BH}$, y por lo tanto también es proporcional al tamaño del agujero negro, definido por su radio de Schwarzschild [24]. Esta proporcionalidad puede expresarse como:  

\begin{equation} 
\lambda \sim R_{S} \approx \dfrac{GM_{BH}}{c^{2}}. 
\end{equation}

Sabemos que un agujero negro radia como un cuerpo negro. De acuerdo con la ley del desplazamiento de Wien [25], la temperatura absoluta $T$ de un cuerpo negro y la longitud de onda $\lambda$ donde se produce el máximo de emisión se relacionan como: 

\begin{equation} 
\lambda T \approx \frac{\hbar c}{k} \sim 10^{-3} m\cdot K. 
\end{equation}

Eliminando $\lambda$ entre las ecuaciones (18) y (19):

\begin{equation} 
T \approx T_{H} \approx \frac{\hbar c^{3}}{kGM_{BH}}. 
\end{equation}

Esta relación y la ecuación (5) de Hawking son formalmente idénticas, y sólo difieren en los factores numéricos adimensionales.\\  

Estimemos ahora el tiempo de evaporación, $t_{ev}$. Para un observador en reposo respecto de un agujero estático, la energía interna $U$ solo puede depender de $M_{BH}$, de modo que $U = M_{BH}c^{2}$. Además, la energía típica de una partícula de la radiación de Hawking es $kT_{H}$. Así, el número total $n$ de partículas que componen el agujero negro puede estimarse como:

\begin{equation} 
n \sim \frac{M_{BH}c^{2}}{kT_{H}}. 
\end{equation}

Por otra parte, como $c$ es la rapidez característica de una partícula de la radiación de Hawking, y $R_{S}$ es la longitud característica, dimensionalmente el tiempo característico que cada partícula demora en ser emitida viene dado por: 

\begin{equation} 
\tau \sim \frac{R_{S}}{c}. 
\end{equation}

En otras palabras, esta expresión establece el intervalo típico que transcurre desde que la partícula es creada hasta que se convierte en real. Si suponemos toscamente que las $n$ partículas son emitidas una por una desde el horizonte, el lapso que un agujero negro demora en radiar todas las partículas que lo componen es:

\begin{equation} 
t_{ev} \sim n\tau \approx \frac{G^{2}M_{BH}^{3}}{c^{4}\hbar} \approx 10^{70}s \left( \frac{M_{BH}}{M_{\odot}}\right)^{3}, 
\end{equation}

donde se han introducido las ecuaciones (1) y (5) para $R_{S}$ y $T_{H}$, respectivamente. Esta ecuación es equivalente a la estimación cruda de Hawking, ecuación (9), y  difiere en un factor $10^{3}$ de la estimación más exacta de Page, ecuación (10). \\ 

Finalmente, estimemos $S_{BH}$. Para ello, vamos a utilizar el hecho que dimensionalmente la entropía es energía por unidad de temperatura. A partir de esta idea, es posible estimar $S_{BH}$ mediante el siguiente  cociente:

\begin{equation} 
S_{BH} \approx \frac{U}{T_{H}} = \frac{M_{BH}c^{2}}{T_{H}} \approx \frac{kGM_{BH}^{2}}{c\hbar}. 
\end{equation}

Resolviendo la ecuación (12) para $M_{BH}^{2}$, introduciendo este valor en la ecuación (24) y utilizando la longitud de Planck, ecuación (13), se obtiene:

\begin{equation} 
S_{BH} \approx \frac{kc^{3}A}{\hbar G} \approx \frac{kA}{l_{P}^{2}}. 
\end{equation}

Esta expresión es formalmente idéntica a las ecuaciones (11) y (14) de Hawking.

\section{¿Existe evidencia de agujeros negros cuánticos?}

Como hemos señalado antes, los efectos cuánticos predichos por Hawking solo son importantes para agujeros negros comparativamente pequeños y poco masivos. En el caso de los agujeros estelares los efectos cuánticos son despreciables, lo cual significa que estos objetos tienen un comportamiento aproximadamente clásico, y pueden describirse satisfactoriamente mediante la relatividad general. Por lo tanto, la expresión \textit{agujeros negros cuánticos} debiera reservarse para los agujeros pequeños y poco masivos. \\ 

La existencia de agujeros cuánticos fue sugerida por Hawking en su artículo seminal de 1974 [4]. En la sección 4 vimos que para esta clase de objetos se tiene que $T_{H} \sim 10^{10}K$, una temperatura muy elevada, y que de acuerdo con la ecuación (19), tiene un máximo de emisión en la región de los rayos gamma, $\lambda \sim 10^{-13}m$, de modo que estos objetos no tienen nada de negros. Por otra parte, si en la ecuación (10) tomamos $M_{BH} \sim 10^{12}kg$, se encuentra que $t_{ev}\sim 10^{17}s$, una cifra que es del orden de la edad del universo, lo cual significa que los agujeros cuánticos se habrían generado poco después del Big Bang, y debieran estar terminando de evaporarse en el presente. Por esta razón también se les conoce como \textit{agujeros negros primordiales}.\\

Desde que Hawking propuso la existencia de los agujeros negros cuánticos, los astrónomos han sondeado el espacio en su búsqueda. No obstante, pese a los esfuerzos desplegados, no existe evidencia sólida de agujeros negros evaporándose o emitiendo radiación de Hawking en forma de rayos gamma [10]. Por otra parte, los agujeros negros menos masivos para los que existe evidencia astronómica son los estelares, cuyo comportamiento cuántico sabemos que es completamente indetectable\footnote{También se ha especulado con la posibilidad de generar agujeros negros cuánticos en grandes aceleradores de partículas como el LHC (Large Hadron Collider), pero tampoco se dispone de evidencia empírica que respalde esta idea.}. ¿Significa esto que los descubrimientos de Hawking están condenados a permanecer en un terreno teórico? La respuesta es no, ya que recientemente se ha abierto un nuevo campo de investigación sin ninguna relación con la astronomía, y que podría permitir la corroboración de las teorías de Hawking.  \\

El primero en proponer que la comprobación empírica de las teorías de Hawking quizá no está en las estrellas, fue el físico teórico canadiense William George Unruh, quien en 1981 sugirió que la temperatura y la radiación de Hawking podrían ser observadas mediante un modelo análogo del horizonte de un agujero negro [26]. Un modelo análogo es un sistema diseñado para imitar el comportamiento de algún objeto o sistema físico difícil de observar en su estado natural. Por lo tanto, el objetivo del modelo análogo es reproducir lo más fielmente el comportamiento físico del fenómeno original para así entender sus propiedades. El año 2014, el físico israelí Jeff Steinhauer del departamento de física del \textit{Israel Institute of Technology} (Technion), logró crear un modelo análogo como el sugerido por Unruh [27]. El modelo análogo del horizonte desarrollado por Steinhauer le permitió observar la emisión de radiación de Hawking, con una temperatura dada aproximadamente por la ecuación (5). Recientemente, Steinhauer y su equipo han logrado extender y mejorar los resultados de su experimento del 2014 [28]. Si bien los resultados obtenidos no son muy exactos debido al ruido inherente al proceso de medición, los resultados nuevamente concuerdan con la predicción de Hawking, al menos dentro de la precisión con la que el experimento permite identificar una temperatura. Aunque los detalles del experimento de Steinhauer son complejos, y su análisis sobrepasa los alcances de este trabajo, lo importante para nuestros propósitos es destacar que quizá la corroboración de las teorías de Hawking no se encuentra en las estrellas sino en la Tierra. Sin embargo, el experimento de Steinhauer no es concluyente y necesita ser confirmado por otros.\\

Pero los resultados de Steinhauer y su equipo no son acontecimientos aislados, ya que forman parte de un activo campo de investigación experimental en gravedad análoga que se encuentra en rápido crecimiento. Sin embargo, de momento los hallazgos en este campo son inciertos debido al ruido experimental, y a la consiguiente dificultad para obtener información robusta. Es de esperar que en la medida que la tecnología avance y las dificultades técnicas puedan ser superadas, la investigación en gravedad análoga aporte grandes descubrimientos a la física de los agujeros negros, especialmente en lo relativo al trabajo de Hawking. El lector interesado puede encontrar información técnica actualizada sobre gravedad análoga y temas afines en [29]. Un artículo no técnico muy recomendable sobre gravedad análoga, agujeros negros y termodinámica es [30]. \\

Vemos entonces que aún queda mucho trabajo por hacer para corroborar empíricamente los hallazgos de Hawking, de modo que en el corto plazo se vislumbra difícil obtener resultados empíricos sólidos, ya sea en gravedad análoga o en astrofísica. En cualquier caso, la importancia de la contribución del genio británico al estudio de los agujeros negros es indudable, y cuando las implicancias físicas de su contribución sean plenamente comprendidas, seguramente será considerada una de las grandes revoluciones científicas el siglo XX.

\section*{Agradecimientos}
Quisiera agradecer a mi amigo y colega, Walter Wussenius, por sus útiles comentarios durante la redacción de este artículo. 

\section*{Referencias}
[1] A. Einstein, Die Grundlage der allgemeinen Relativitätstheorie, Annalen der Physik, 354 (1916) 769-822.

\vspace{2mm}

[2] A. Einstein, The Collected Papers of Albert Einstein, Princeton University Press, Princeton, 1997.

\vspace{2mm}

[3] K.S. Thorne, Agujeros negros y tiempo curvo: El escandaloso legado de Einstein, Crítica, Barcelona, 2000.

\vspace{2mm}

[4] S.W. Hawking, Black Hole explosions?, Nature, 248 (1974) 30-31.

\vspace{2mm}

[5] S.W. Hawking, Particle creation by black holes, Communications in Mathematical Physics, 43 (1975) 199-220.

\vspace{2mm}

[6] M.C. LoPresto, Some Simple Black Hole Thermodynamics, The Physics Teacher, 41 (2003) 299-301.

\vspace{2mm}

[7] J. Pinochet, The Hawking temperature, the uncertainty principle and quantum black holes, Physics Education, 53 (2018) 1-6.

\vspace{2mm}

[8] J.A. Wheeler, R. Ruffini, Introducing the black hole, Physics Today, (1971) 30-41.

\vspace{2mm}

[9] V.P. Frolov, I.D. Novikov, Black Hole Physics: Basic Concepts and New Developments, Springer Science, Denver, 1998.

\vspace{2mm}

[10] V.P. Frolov, A. Zelnikov, Introduction to Black Hole Physics, Oxford University Press, Oxford, 2011.

\vspace{2mm}

[11] S.W. Hawking, Historia del tiempo: Del big bang a los agujeros negros, 3 ed., Crítica, Barcelona, 1988.

\vspace{2mm}

[12] S.W. Hawking, A brief history of time, Bantam Books, New York, 1998.

\vspace{2mm}

[13] S.B. Giddings, Hawking radiation, the Stefan–Boltzmann law, and unitarization, Physics Letters B, 754 (2016) 39-42.

\vspace{2mm}
[14] S.W. Hawking, Black Holes in General Relativity, Communications in Mathematical Physics, 25 (1972) 152-166.

\vspace{2mm}

[15] D. Christodoulou, Reversible and Irreversible Transformations in Black-Hole physics, Physical Review Letters, 25 (1970) 1596-1597.

\vspace{2mm}

[16] D. Christodoulou, Reversible Transformations of a Charged Black Hole, Physical Review D, 4 (1971) 3552-3555.

\vspace{2mm}

[17] R. Penrose, R.M. Floyd, Extraction of Rotational Energy from a Black Hole, Nature, 229 (1971) 177-179.

\vspace{2mm}

[18] B.J. Carr, S.B. Giddings, Quantum black holes, Scientific American, (2005) 48-55.

\vspace{2mm}

[19] R. J. Adler, The Generalized Uncertainty Principle and Black Hole Remnants, General Relativity and Gravitation 33 (2001) 2101-2108.

\vspace{2mm}

[20] D.N. Page, Particle emission rates from a black hole: Massless particles from an uncharged, nonrotating hole, Physical Review D, 13 (1976) 198-206.

\vspace{2mm}

[21] J.D. Bekenstein, Black Holes and Entropy, Physical Review D, 7 (1973) 2333-2346.

\vspace{2mm}

[22] R.J. Adler, Six easy roads to the Planck scale, American Journal of Physics, 78 (2010) 925-932.

\vspace{2mm}

[23] S.W. Hawking, Black Holes and Thermodynamics, Physical Review D, 13 (1976) 191-197.

\vspace{2mm}

[24] N.B. Birrel, P.C.W. Davies, Quantum Fields in Curved Spaces, Cambridge University Press, Cambridge, 1982.

\vspace{2mm}

[25] P.A. Tipler, R.A. Llewellyn, Modern Physics, 6 ed., W. H. Freeman and Company, New York, 2012.

\vspace{2mm}

[26] W.G. Unruh, Experimental Black-Hole Evaporation?, Physical Review Letters, 46 (1981) 1351-1353.

\vspace{2mm}

[27] J. Steinhauer, Observation of self-amplifying Hawking radiation in an analogue black-hole laser, Nature Physics, 10 (2014) 864-869.

\vspace{2mm}

[28] J. R. Muñoz de Nova, K. Golubkov, V. I. Kolobov, J. Steinhauer, Observation of thermal Hawking radiation at the Hawking temperature in an analogue black hole, arXiv:1809.00913v2 [gr-qc] (2018).

\vspace{2mm}

[29] https://hyperspace.uni-frankfurt.de/2011/05/27/living-reviews-in-relativity-analogue-gravity-and-the-einstein-vlasov-systemkinetic-theory-updates/ [consulta: 28/01/2019].

\vspace{2mm}

[30] C. Barceló, Agujeros negros, sus análogos y termodinámica, Revista Española de Física, 29 (2015) 34-39.

\section*{Enumeración de figuras}

FIGURA 1. Un agujero negro estático tiene una singularidad central ($S$) rodeada por un horizonte esférico cerrado ($H$).\\

FIGURA 2. Un par de partículas virtuales, compuestas por un electrón $(e^{-})$ y un positrón $(e^{+})$ son creados en un determinado instante. En la medida que transcurre el tiempo, las partículas se separan cierta distancia para luego volver a encontrarse, aniquilarse, y ser reabsorbidas por el vacío.\\

FIGURA 3. Cuando se produce un par partícula-antipartícula fuera del horizonte de un agujero negro, una partícula ($e^{+}$) es atrapada por la gravedad y cae dentro del horizonte. La otra partícula ($e^{-}$) logra escapar. Externamente, esta partícula es detectada como radiación de Hawking con temperatura $T_{H}$.  

\end{document}